\documentstyle[preprint,aps]{revtex}

\begin{document}
\draft
\title{Temperature and Pressure Effects on the Resistivity of \\
the Manganese Oxides}
\author{Hae-Young Kee and Jongbae Hong}
\address{Department of Physics Education, Seoul National University,
Seoul 151-742, Korea}
\maketitle

\begin{abstract}
The temperature and pressure effects on the resistivity of the manganese oxides
are studied analytically via the Kondo lattice model with a strong Hund's
coupling.  We obtain analytically the single-particle density of states 
on the Bethe lattice in the large connectivity limit using the 
analytical variant of the dynamic Lanczos method. 
From the density of states for the doped system,
we obtain resistivity and show resistivity drop 
when temperature crosses the magnetic transition point. 
We also demonstrate the effect of pressure
both below and above the transition temperature. 
\end{abstract}

\section{Introduction}
Though manganese oxides had been studied extensively during 50's 
and 60's,\cite{jonker} 
the hole-doped manganese
oxides $La_{1-x}A_xMnO_3 (A=Ca, Ba, Pb, \mbox{or} \ Sr)$ have recently 
driven a considerable attention due to not only the phenomenon of giant
magnetoresistance (GMR)\cite{rmk} but also various interesting
physical properties such as doping induced metal-insulator transition,
antiferromagnetic-ferromagnetic transition, and the structural phase
transition\cite{asami}. These phenomena have been partly explained by the
double exchange mechanism.\cite{jonker,zener} 
The ferromagnetic Kondo lattice model (KLM) due to
strong Hund's coupling\cite{kubo} has been proposed 
to describe the properties of
manganese oxides.
Recently, a mean field theory, which is valid in infinite 
dimensions,\cite{inf} 
has been applied to the KLM in the limit of infinite magnitude of 
local spin.\cite{fu} 
However, the large resistance drop measured in experiment\cite{rmk} 
has not been shown successfully.  
Millis {\it et al.} have recently argued that the strong 
electron-phonon coupling is necessary to explain the colossal
magnetoresistance\cite{millis}.
a new experiment\cite{mich} shows that there is no primary
lattice distortion anomaly at the magnetic transition temperature.

Leaving aside the question how much the polaron physics affects 
the properties of manganese oxides, we are interested in how well 
the KLM with a strong Hund's coupling
describes the resistivity drop upon variation of temperature\cite{schi} and
pressure\cite{jjn,arn,morito} in the manganese oxides. 
We feel that proper calculation of the resistivity in this model is 
necessary to understand how much the electron-phonon interaction 
is important for the explanation of the experiments.  
Even though magnetovolume
contraction\cite{jmde,mri} accompanies across the magnetic transition, 
we do not consider this effect explicitly in this work. 
We suppose that the effect of pressure is to enhance not only hopping 
between conduction electrons but also scattering rate.
These two effects appear differently in ferromagnetic and paramagnetic states.
The purpose of this work is to show quantitatively how much the 
resistivity drops due to magnetic ordering and pressure via
temperature and transfer integral, respectively.

Our work starts from the single-particle density of states (DOS) which has
not  been previously obtained through a rigorous calculation for the KLM. 
We first obtain analytically
the DOS of the KLM on the Bethe lattice in the large connectivity. 
The method we use here, the so called analytic version of the dynamic 
Lanczos method, has proved useful in describing the optical transport 
properties of $V_2O_3$\cite{jongbae}
and the doped cuprates\cite{kee}. In a previous work\cite{ho}, 
analytical calculation of DOS through this method has given 
transparent understanding of the metal-insulator transition 
in the large-dimensional half-filled Hubbard model .

This paper is composed as follows.
In Sec. II, we briefly introduce the formalism, and show the 
single-particle DOS in Sec. III. 
We present the resistivity depending on temperature and
pressure in Sec. IV, and finally give conclusion in Sec. V.

\section{Formalism}
The single-particle DOS, $N_{\sigma}(\omega)$, is given by\cite{ho,fulde}
\begin{eqnarray}
N_{\sigma}(\omega)&=&\frac{1}{\pi N}\lim_{\eta\rightarrow 0^+}\sum_j
{\rm Re}\int_0^{\infty}\langle\{c_{j\sigma}(t),c_{j\sigma}^{\dagger}\}\rangle
e^{i\omega t-\eta t}dt \nonumber \\
                              &\equiv&\frac{1}{\pi}\lim_{\eta\rightarrow
0^+}{\rm Re} \; a_0(z)|_{z=-i\omega+\eta},
\end{eqnarray}
where the angular brackets denote the thermal average and $a_0(z)$, which is
the Laplace transform of
$\langle\{c_{j\sigma}(t),c_{j\sigma}^{\dagger}\}\rangle$.
One notes that $a_0(z)$ corresponds to the on-site Green function and is the
key quantity giving the DOS.

One can observe that $a_0(z)$ is just the projection of $c_{j\sigma}(z)$ onto
$c_{j\sigma}$ in a Liouville space defined by the inner product
$(A,B)=\langle\{A,B^{\dagger}\}\rangle$, where $A$ and $B$ are operators of
the Liouville space, and $B^{\dagger}$ is the adjoint of $B$. This projection
can be obtained most easily in the orthogonalized space. Therefore, we first
construct an appropriate orthogonal space. We choose the first vector $f_0$ as
$f_0=c_{j\sigma}$ and the second using the Gram-Schmidt process with a
linearly independent vector $iLf_0$ where $L$ is the Liouville operator. The
rest orthogonal vectors are obtained via a similar process which may be
expressed by the recurrence relation,
\begin{equation}
f_{\nu+1}=iLf_{\nu}-\alpha_{\nu}f_{\nu}+\Delta_{\nu}f_{\nu-1},
\end{equation}
where $f_{-1}\equiv 0$ and $\Delta_{\nu}=(f_{\nu},f_{\nu})/(f_{\nu-1},f_{\nu-
1})$ for $\nu\geq 1$. Use of the relation $(iLA,B)=-(A,iLB)$ has been made.
Applying Eq. (2) to $f_0(t)=\sum_{\nu=0}^{\infty}a_{\nu}(t)f_{\nu}$, the
projection $a_0(z)$ is represented by the infinite
continued fraction\cite{hong,fulde,mori,zwan,lee}
\begin{equation}
a_0(z)=\frac{1}{z-\alpha_0+\frac{\Delta_1}{z-
\alpha_1+\frac{\Delta_2}{z-\alpha_2+\ddots}}},
\end{equation}
where $\alpha_{\nu}=(iLf_{\nu},f_{\nu})/(f_{\nu},f_{\nu})$.

For a nontrivial system, however, the fractional forms of $\alpha_{\nu}$ and
$\Delta_{\nu}$ require expansions in performing the Gram-Schmidt process
analytically, and allow us an orthogonal set satisfying Eq. (2) exactly only
in an asymptotic regime, e.g., strongly or weakly correlated regime. Since
$\alpha_{\nu}$ provides with energy standard and can be controlled by chemical
potential, $\alpha_{\nu}$ usually has a simple form.  What makes things hard
is $\Delta_{\nu}$ which is the ratio of norms of consecutive bases. Since norm
is usually given by a polynomial composed of positive terms of an expansion
parameter, the ratio of leading terms gives quite good approximation when the
terms of each norm show a similar behavior. This is actually the case we meet
in practice.  This maybe happens since we get the bases through a simple
recurrence relation. We argue that the valid region of the asymptotic dynamics
extends to the regime where the expansion is allowed.  One can recognize it
from the result. One of us
and Lee\cite{hong82,hong93} have shown the dynamics of electron gas in the
asymptotic regimes such as small momentum transfer and large momentum transfer
limit. They have shown that the former is good for the regime $k<k_F$
\cite{hong82}, where $k$ and $k_F$ are momentum transfer 
and Fermi momentum, respectively and the
latter agrees well with experiment for the momentum transfer 
$k\approx 2k_F$\cite{hong93}.

\section{Single-Particle Density of State}
We now apply this formalism to the KLM which is written as
\begin{equation}
H=-\sum_{\langle i,j \rangle ,\sigma} t_{ij} (c^+_{i,\sigma}c_{j,\sigma}+h.c.)
   -J\sum_i {\bf S}_i\cdot {\bf \sigma}_i,
\end{equation}
where angular bracket means nearest neighbor.
The first term represents hopping processes of the itinerant electrons
in $e_g$ orbital and the second represents ferromagnetic Kondo coupling between 
localized spin ($S=3/2$) and itinerant electron spin on $Mn^{3+}$ ion 
coming from $t_{2g}$ and $e_g$ orbitals, respectively. 
The electron spin ${\bf \sigma}_i$ 
can be represented in terms of fermion operators.
Therefore we have the following Hamiltonian.
\begin{equation}
H=-\sum_{\langle i,j \rangle ,\sigma} t_{ij}
   (c^+_{i,\sigma}c_{j,\sigma}+h.c.)
   -\frac{J}{2}\sum_i [ S^{-}_i c_{i\uparrow}^{\dagger}c_{i\downarrow}
 +S^{+}_ic_{i\downarrow}^{\dagger}c_{i\uparrow}
 +S^z_i (c_{i\uparrow}^{\dagger}c_{i\uparrow}
     -c_{i\downarrow}^{\dagger}c_{i\downarrow})]
\end{equation}
where ${\bf S}= (S^x,S^y,S^z)$ is a classical spin.

The key part of this work is to construct the orthogonal bases $f_{\nu}$'s
along with $\alpha_{\nu}$ and $\Delta_{\nu}$.
Using the recurrence relation (2) with $f_0=c_{j\uparrow}$ and considering
$J>>t_{ij}$ for expansion, we obtain the orthogonal bases 
by taking the leading terms as follows:
\begin{eqnarray}
f_{4\nu}&=&  J^{2\nu}\overbrace{\Sigma_j^{\prime}\cdots
     \Sigma_q^{\prime}}^{2\nu}\overbrace{t_{kj}\cdots t_{pq}}^{2\nu}
   \overbrace{{\bf T}_k\cdots {\bf T}_p}^{2\nu}
    c_{q \uparrow}\nonumber\\
f_{4\nu+1} &=& i \frac{J^{2\nu+1}}{2} \overbrace{\Sigma_j^{\prime}\cdots
 \Sigma_q^{\prime}}^{2\nu}\overbrace{t_{kj}\cdots t_{pq}}^{2\nu}
       \overbrace{{\bf T}_k\cdots {\bf T}_p}^{2\nu}
   ( \Delta S^z_q c_{q\uparrow}+  S^{-}_q c_{q\downarrow})
       \nonumber\\
f_{4\nu+2} &=& - J^{2\nu+1}\overbrace{\Sigma_j^{\prime}\cdots
\Sigma_r^{\prime}}^{2\nu+1}\overbrace{t_{kj}\cdots t_{qr}}^{2\nu+1}
     \overbrace{{\bf T}_k\cdots {\bf T}_p}^{2\nu}
  [(S^z_q+S^z_r) c_{r\uparrow}+(S^-_q+S^-_r) c_{r\downarrow}]\nonumber\\
f_{4\nu+3} &=& -i \frac{J^{2\nu+2}}{2} \overbrace{\Sigma_j^{\prime}\cdots
   \Sigma_r^{\prime}}^{2\nu+1}\overbrace{t_{kj}\cdots t_{qr}}^{2\nu+1}
 \overbrace{{\bf T}_k\cdots {\bf T}_p}^{2\nu}\nonumber \\
  & & \hspace{5mm} \times [(S^z_q+ S^z_r) \Delta S^z_r c_{r\uparrow}
   +(S^-_q+S^-_r)S^{+}_r c_{r\uparrow}
   +(S^z_q S^z_r-S^-_q S^z_r) c_{r\downarrow} \nonumber\\
  & & \hspace{5mm} -(S^-_q + S^-_r) \langle S^z \rangle c_{r\downarrow}]
\end{eqnarray}                                                     
where $\nu\geq 0$,
$\Delta S^z =S^z-\langle S^z \rangle $, $\langle S^z_k \rangle =\langle
S^z_j \rangle \equiv \langle S^z \rangle$, and ${\bf T}_k={\bf S}_k+{\bf S}_j$.
We consider only nearest neighbor hopping so that
$j$ is the nearest neighbor of $k$, $l$ is the nearest neighbor of $j$,
and  the prime in summation means $k \neq j \neq \cdots \neq r$.
%We introduce  $\{ {\bf T}_k \}$ as a notation
%to simplify the forms of $f_{\nu}$'s which are the multiples of
%$\{ {\bf T}_k \}$ and $c_{\sigma}$.
The inner operation of  $\{ {\bf T}_k \}$ product is defined as the
scalar product only with itself.
This operation is valid for classical spins.
We obtain the inner products as follows
\begin{eqnarray}
(f_1,f_1)  &=& \langle \frac{J^2}{4} [(\Delta S^z_k)^2+S^-_k S^{+}_k] \rangle
\nonumber\\
 &=&  \frac{J^2}{4} (\langle {\bf S}^2_k \rangle- \langle {\bf S}_k \rangle^2)
 \nonumber\\
(f_2,f_2)  &=& \langle J^2 \Sigma_j^{\prime} t_{kj}^2 [(S^z_k+S^z_j)^2
  +(S^-_k+S^-_j)(S^{\dagger}_k+S^{\dagger}_j)] \rangle \nonumber\\
 &= &  J^2 \Sigma_j^{\prime} t_{kj}^2 \langle ({\bf S}_k+{\bf S}_j)^2
 \rangle \nonumber\\
 &= &  J^2 \Sigma_j^{\prime} t_{kj}^2 \langle {\bf T}_k^2 \rangle\nonumber\\
(f_3,f_3) &=& \frac{J^4}{4}
\Sigma_j^{\prime} t_{kj}^2 \langle {\bf T}_k^2 \rangle
  (\langle {\bf S}^2_j \rangle- \langle {\bf S}_j \rangle^2) \nonumber\\
(f_4,f_4) &=& J^4 \Sigma_j^{\prime}\Sigma_l^{\prime} t_{kj}^2 t_{jl}^2
  \langle {\bf T}_k^2 {\bf T}_j^2  \rangle  \nonumber\\
 & & \hspace{2cm} \vdots
\end{eqnarray}
These bases (6) satisfy the orthogonal condition as follow.
\begin{eqnarray}
(f_{4 \nu}, f_{4 \nu+1})&=&\langle   \frac{J^{4\nu+1}}{2}
    \overbrace{\Sigma_j^{\prime}\cdots\Sigma_q^{\prime}}^{2\nu}
    \overbrace{t_{kj}^2\cdots t_{pq}^2}^{2\nu}
    \overbrace{{\bf T}_k^2\cdots {\bf T}_p^2}^{2\nu}[
    ( S^z_q -\langle  S^z \rangle) \{c_{q\uparrow}, c_{q\uparrow}^{\dagger}\}
     +S^-_q \{c_{q\downarrow}, c_{q\uparrow}^{\dagger} \}] \rangle \nonumber\\
      &=&  \frac{J^{4\nu+1}}{2}
      \overbrace{\Sigma_j^{\prime}\cdots\Sigma_q^{\prime}}^{2\nu}
      \overbrace{t_{kj}^2\cdots t_{pq}^2}^{2\nu}
      \overbrace{\langle {\bf T}_k^2 \rangle \cdots
      \langle {\bf T}_p^2 \rangle}^{2\nu} \langle ( S^z_q-\langle  S^z \rangle)
      \rangle = 0
\end{eqnarray}
and so on.
We take the approximation neglecting the spatial correlations between
the localized spins of $Mn^{3+}$
, that is,
$\langle {\bf S}_k\cdot {\bf S}_j\rangle=\langle {\bf S}_k\rangle
\cdot \langle {\bf S}_j\rangle \equiv \langle {\bf S}\rangle^2$
for $k \neq j$.
where $\langle {\bf S}\rangle$ is independent of sites.
Therefore $\langle {\bf T}^2 \rangle =2 \langle {\bf S}^2 \rangle
+2 \langle {\bf S} \rangle^2$.
Note that this mean field approximation (MFA) is valid in the limit of
large connectivity.
Within this approximation, one obtains $\alpha_{\nu}$ and $\Delta_{\nu}$
as follows:
\begin{eqnarray}
\alpha_{2\nu}&=&i\frac{J}{2}\langle {\bf S} \rangle
      \equiv i\alpha,\nonumber\\
\alpha_{2\nu+1} &= &-i\frac{J}{2} \langle {\bf  S} \rangle
      \equiv -i\alpha,\nonumber\\
\Delta_{2\nu+1} &= &\frac{J^2}{4}
(\langle {\bf S}^2\rangle-\langle {\bf S} \rangle^2)
	 \equiv \Delta_o, \nonumber\\
\Delta_{2\nu+2} &=& 4 t^2_*
    \frac{ (\langle {\bf S}^2\rangle+\langle {\bf S} \rangle^2)}
          {(\langle {\bf S}^2\rangle-\langle {\bf S} \rangle^2)}
	\equiv \Delta_e,
\end{eqnarray}
for $\nu\geq 0$, where $t_*=\frac{t}{\sqrt{2q}}$
and $q$ is the coordination number which is taken to be very large.

Now the DOS of the up-spin conduction electron at site $j$ is given by
Eqs. (1), (3), and (9), i.e.,
\begin{equation}
N_{\uparrow}(\omega)=\frac{1}{\pi}
       \frac{\sqrt{4\Delta_e(\omega^2 - \alpha^2)
    -(\Delta_o-\Delta_e-(\omega^2 - \alpha^2))^2}}
    {2|\omega - \alpha|\Delta_e}
\end{equation}
This DOS shows two separated bands of width $W=
\sqrt{\alpha^2+(\sqrt{\Delta_o} +\sqrt{\Delta_e})^2}
-\sqrt{\alpha^2+(\sqrt{\Delta_o} -\sqrt{\Delta_e})^2}$, 
which is $4 t_*$ when $\langle {\bf S} \rangle =0$, i.e., in 
paramagnetic state.
Eqs. (9) and (10) show that the DOS depends on the magnetic properties 
of the system through $\langle{\bf S}\rangle$ and $\langle{\bf S}^2\rangle$ of
the localized spin. 

Below the transition temperature $T_c$, the system is ferromagnetic and the
expectation value of  $\langle {\bf S}\rangle$ can be obtained from the
expermental data and $\langle {\bf S}^2 \rangle$ is nothing but $S(S+1)$.
Above $T_c$, where the
system is paramagnetic, $\langle{\bf S}\rangle$ becomes $0$
while $\langle{\bf S}^2 \rangle$ remains $S(S+1)$.

Fig. 1 shows the DOS for the Hund's coupling strength $J/W=4$ and a doping
concentration $x=0.3$ using  $\langle {\bf S} \rangle=0$, $0.4$, and $0.8$ with
dotted, dashed, and solid lines, respectively.
The vertical lines indicate
positions of the chemical potential.
We only show the lower band since the upper band is nothing but point 
reflection of lower band about zero frequency, and
does not affect resistivity within the range of
temperature in which we are interested.
The DOS clearly shows the shift of the spectral weight to lower energy
side as local spins are ordered. 
Magnetic ordering enhances up-spin spectral weight and
diminishes down-spin spectral weight remarkably.
Note that the double exchange model
leads us to the rigid band picture 
and the position of chemical potential is determined 
only by the amount of filling which results in 
the change of resistivity upon doping concentration, though     
our DOS, Eq. (10), does not contain the carrier concentration 
$\langle n\rangle$.  \cite{note}
Our DOS satisfies sum rule exactly.

\section{Resistivity}
The physical quantity we want to examine in this work 
is resistivity affected by ordering of magnetic moments and pressure. 
We show specifically in
what follows how magnetic ordering and pressure affect resistivity via
obtaining optical conductivity.

Optical conductivity in infinite dimensions is obtained using the
following formula\cite{th}.
\begin{equation}
\sigma(\omega)=\sigma_0 t^2_* \int d w^{\prime} \int d\epsilon A_0(\epsilon)
 A(\epsilon,\omega^{\prime}) A(\epsilon,\omega^{\prime}+\omega)
  \frac{f(\omega^{\prime})-f(\omega^{\prime}+\omega)}{\omega}
       \equiv \sigma_0 I(\omega),
\end{equation}
where $A_0(\epsilon)=\sqrt{2-\frac{\epsilon^2}{t^2_*}}/t_* \pi$
for the Bethe lattice, $f(\omega)$ is the Fermi distribution function, and
$\sigma_0=\frac{\pi e^2 a^2 N}{2 \hbar V}$ where $a, N, V$ are lattice
constant, number of lattice sites, and volume, respectively.

One can get $A(\epsilon,\omega)$ by calculating the self-energy
$\Sigma(\omega)$ using the relation $\Sigma_{\sigma}(\omega)=\omega-
\frac{t_*^2G_{\sigma}(\omega)}{2}-\frac{1}{G_{\sigma}(\omega)}$ applicable to
the Bethe lattice\cite{jongbae,gk} and $G(\omega)=-ia_0(-i\omega)$. The dc
resistivity is obtained by taking the zero-frequency limit 
for the inverse optical
conductivity, i.e.,
\begin{equation}
\rho(T)=\lim_{\omega\rightarrow
0}\frac{1}{\sigma(\omega,T)}=\frac{1}{\sigma_0I(0,T)},
\end{equation}
The unit of conductivity is provided by
$\sigma_0$ which, in three dimensions, approximately gives
Mott's minimum metallic conductivity 
$\sigma_{Mott} \sim 10^3 (\Omega cm)^{-1}$.\cite{fu,th} 
However the absolute value of the resistivity should depend on
the microscopic parameters such as lattice constant in $\sigma_0$,
so one cannot determine $\sigma_0$ precisely.
It usually varies of $O(1)$ depending on sample.

Fig. 2 shows temperature dependences of the resistivity for
$J/W=4$ with $x=0.3$ 
and magnetization around transition temperature. 
The magnetization is obtained as
${\bf M}=g  [\langle {\bf S} \rangle+ \frac{1}{2}
(\langle n_{\uparrow} \rangle-\langle n_{\downarrow}\rangle)]$,
where $g$ is the $g$-factor.
The temperature dependence of $\langle {\bf S} \rangle$ has been conjectured
as $\langle {\bf S} \rangle =
\sqrt{\langle {\bf S}^2 \rangle (1-\frac{T}{T_c})}$
from the mean-field type result, i.e., 
$\langle {\bf S} \rangle \propto \sqrt{1-\frac{T}{T_c}}$.
This magnetic ordering curve shown in Fig. 2 
is remarkablely similar to the experimental data 
of magnetization\cite{mich}. 
It shows that the larger magnetic ordering induces the larger 
resistivity drop.
Since it has been known that $La_{1-x}Sr_x
MnO_3$ has a bandwidth $W\sim 1eV$\cite{morito} ,
we set $t_*=0.25 eV$ so that the bandwidth $W= 4t_*=1 eV$
in the paramagnetic state, and $J=4 W=4 eV$ 
which is known as a reasonable value of Hund's coupling.\cite{fu}  
The transition temperature is set $T^0_c=0.13t_*=377 K$
to compare with the experiment\cite{mich}.

We try to understand resistivity drop more physically through 
the changes of scattering rate and bandwidth versus local spin fluctuation. 
In Fig. 3, we show the change of scattering rate, $1/\tau_{\uparrow}$,
and bandwidth across $T_c$
to see what governs the resistivity drop.
The scattering rate is given by
$1/\tau_{\uparrow} = 2 Im \Sigma_{\uparrow} (\omega=\mu)$,
where $\mu$ is chemical potential.
In Eq. (9) and (10), we see that the DOS has a local spin fluctuation term 
$\langle (\Delta {\bf S})^2 \rangle$.
The local spin fluctuation decreases as magnetic ordering increases,
which results in broadening of bandwidth and decreasing of
scattering rate of up-spin conduction electron which is 
parallel to the localized spin.
The scattering rate of down-spin conduction electron increases 
quite rapidly as local spins are ordered. 
Fig. 3 demonstrates that more spin-disorder scattering occurs
in the paramagnetic state, and the spin-disorder strongly  
affects the scattering rate governing the resistivity.

We show pressure effect by changing the transfer integral 
$t_*$ for a fixed Hund's coupling strength $J$.
Though the pressure affects both $J$ and bandwidth $W$,
its effect is much larger for $W$ than for $J$.
Therefore we make $J$ fix and $W$ increase as pressure increases.
The proportionality  $T_c \propto W$ is valid under MFA\cite{kubo,fu},
so that $T_c$ is  taken at $414 K$ for $W=1.1 eV$
compared with $377 K$ for $W=1.0 eV$.
Fig. 4 clearly shows that no pressure effect appears 
in the resistivity above $T_c$.
The origin of this result can be understood as follows.
%comes from the fact that the pressure effect 
%is exactly canceled out the increased  scattering rate above $T_c$.
We obtain the effective mass using 
$m^*=1-\frac{\partial Re\Sigma}{\partial \omega}|_{\omega=\mu}$.
The  effective mass decreases by increasing $t^*$, while
the scattering rate increases nearly the same amount by increasing $t^*$,
so the effect of change of $t^*$ in the resistivity disappears
in the paramagnetic state,
if we consider $\rho \propto m^*/\tau $ with assuming constant number of
charge carriers.
Therefore we have to consider the lattice constant appearing in $\sigma_0$,
which will be changed under pressure, to explain 
the experimental results showing remarkable resistivity drop
even above $T_c$ under pressure.
The lattice dynamics may play an important role in the paramagnetic regime 
near $T_c$.\cite{millis}

\section{Conclusion}
We study the temperature and pressure effects on the
resistivity of the hole doped manganese oxides by calculating the DOS and
optical conductivity for the KLM with strong Hund's coupling using the Lanczos
continued fraction formalism along with MFA. 
These two effects appear in both  DOS and  resistivity, and the results
show  that the smaller spin disorder induces the smaller resistivity. 
Our result also shows that there is no resistivity drop in paramagnetic state,
above $T_c$, under pressure, 
which has a clear discrepancy with many experimental
data\cite{jjn,arn,morito}.
It demonstrates that either double exchange model losts the main physics 
in the GMR phenomenon
or treating the double exchange model within MFA yields too simple physics,
which shows that only local spin fluctuation can change resistivity.
Above and near  $T_c$, however,  the magnetovolume effect may cause further
increasing of resistivity due to decreasing of lattice constant in
$\sigma_0$ of Eq. (12).
This effect should be included to improve the resistivity drop 
just above $T_c$
and to show the pressure effect shown in experiments\cite{jjn,arn,morito},
to some extent.
Thus present results indicate that lattice distortion may 
be crucial as far as the resistivity in paramagnet is concerned
and polaron may play an important role in this sense.

\acknowledgements
%\centerline{ACKNOWLEDGEMENT}

This work  has   been supported by SNU-CTP,  Basic  Science  Research
Institute  Program,  Ministry   of Education and the Daewoo Fundation.
One of authors (H-Y Kee) appreciates Y-B Kim and P. Coleman for
useful discussion.

\noindent
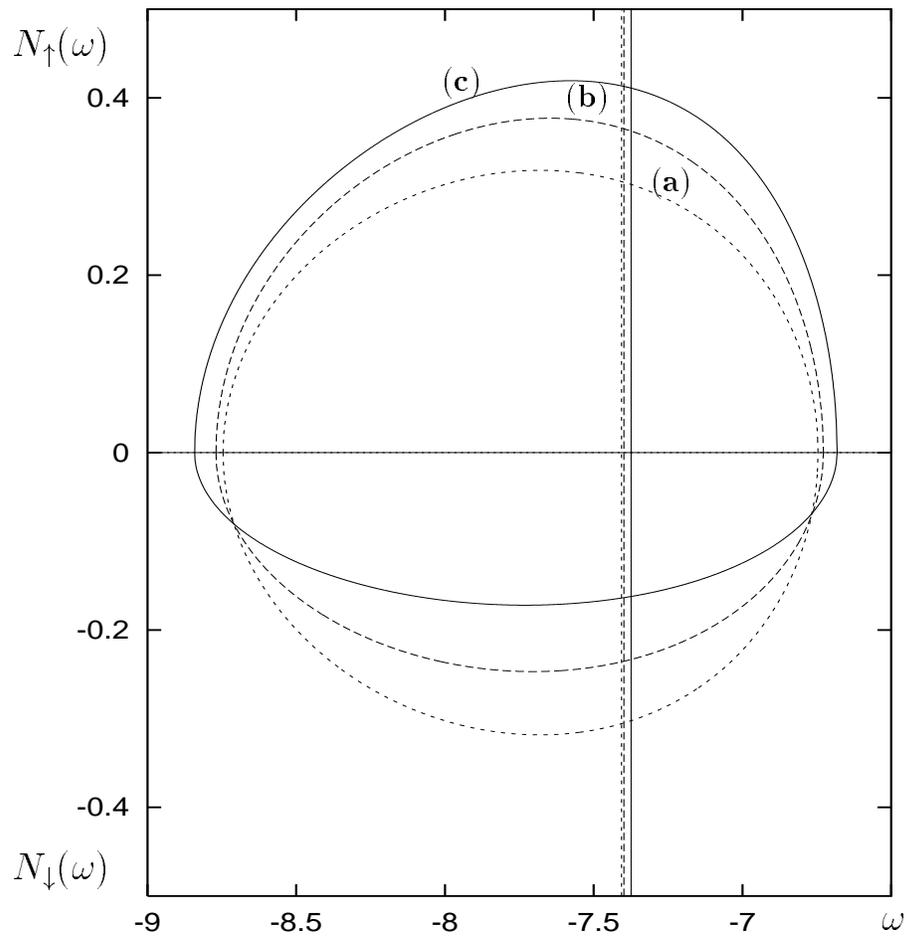
\begin{figure}
\caption{ DOS when $J/W=4$ for paramagnetic state (a) with dotted line and
ferromagnetic state (b) with dashed line  and (c) with solid line. We use
$\langle {\bf S} \rangle = 0.4$ (b) and $0.8$ (c).
The vertical lines indicate the position of the chemical potential for $x=0.3$.}
\end{figure}

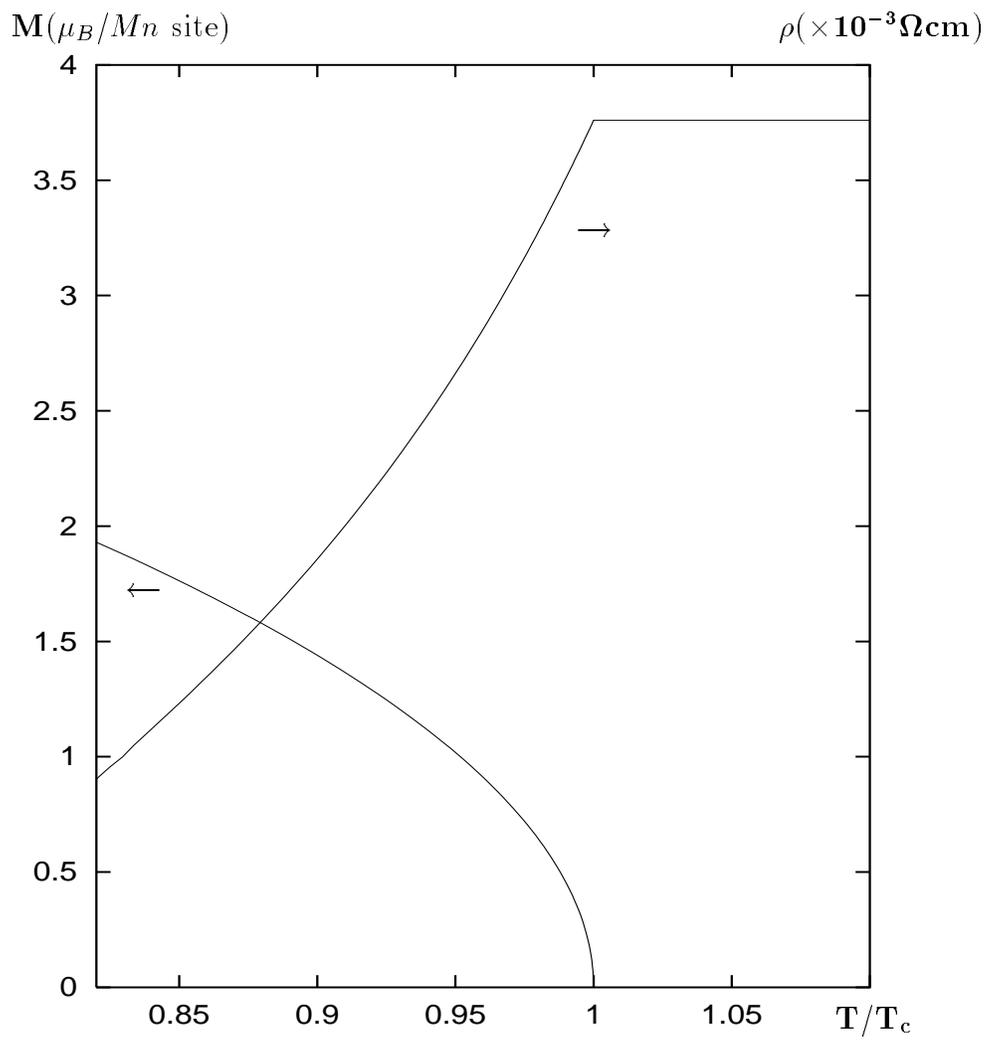
\begin{figure}
\caption{Resistivity and magnetization
for $J/W=4$ and $x=0.3$
are shown around the transition temperature.
Magnetization has been  obtained as 
$g  [\langle {\bf S} \rangle+ \frac{1}{2}
(\langle n_{\uparrow} \rangle-\langle n_{\downarrow}\rangle)]$, 
where $g$ is the $g$-factor. }
\end{figure}

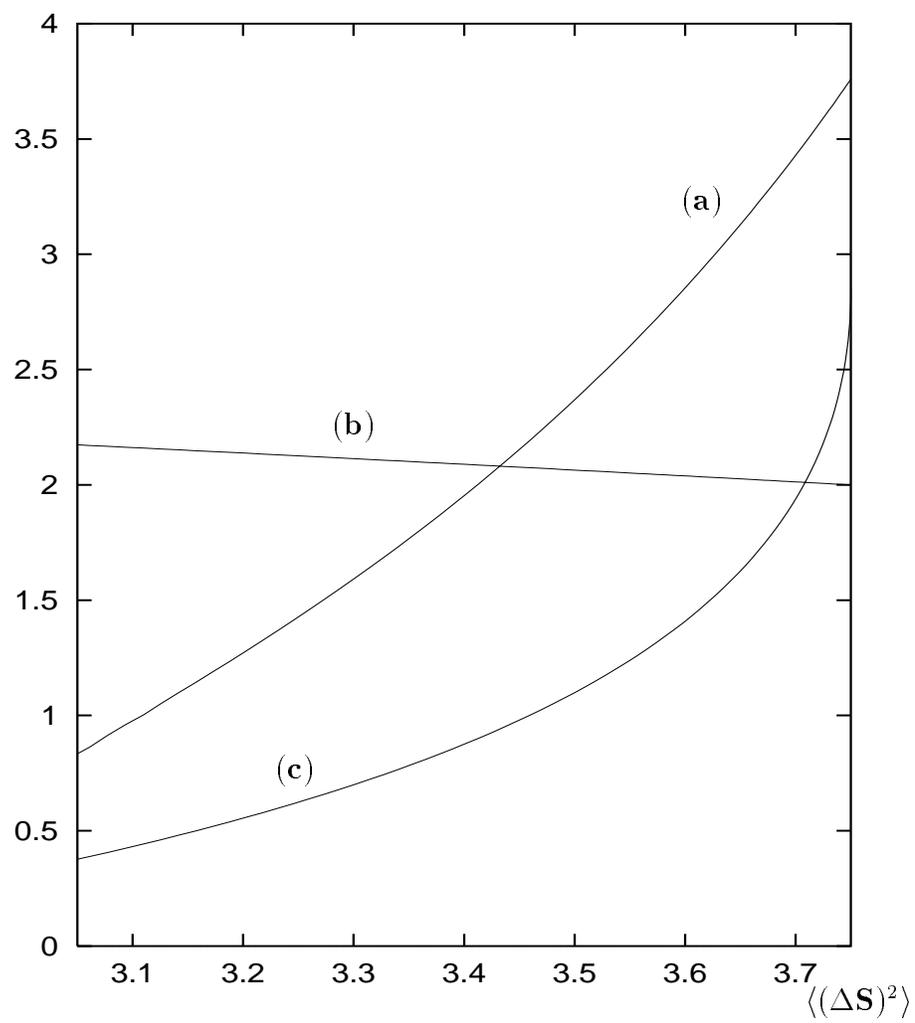
\begin{figure}
\caption{ Resistivity (a),
  bandwidth (b) and scattering rate (c) in units of $10^{-3}\Omega$ cm,
$2 t^*$, and\\
$\frac{1}{6}$ sec.$^{-1}$ respectively
  v.s. local spin fluctuation are drawn.}
\end{figure}

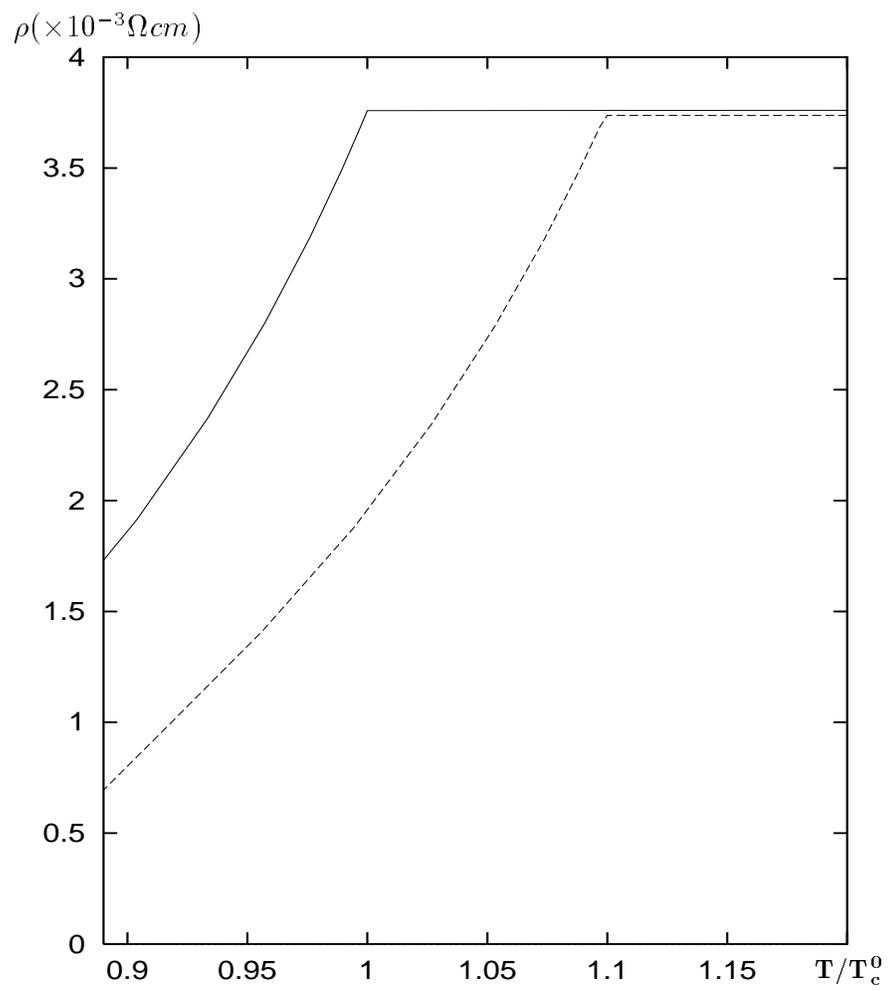
\begin{figure}
\caption{Resistivities for $W=1.0 eV$(solid line) and $1.1 eV$(dashed line)
with $J=4 eV$, $x=0.3$ are shown around the transition temperature, where
$T_c^0$ denotes  the transition temperature for $W=1.0 eV$}
\end{figure}

\begin{thebibliography}{99}
\bibitem{note} The filling factor dependence in the DOS appears explicitly
when the on-site Coulomb repulsion is included in the Hamiltonian.
\end{thebibliography}
\end{document}